\documentclass[12pt]{iopart}

\expandafter\let\csname equation*\endcsname\relax
\expandafter\let\csname endequation*\endcsname\relax

\usepackage{amsfonts}
\usepackage{amsmath}
\usepackage{amssymb}
\usepackage{bm}
\usepackage{dcolumn}
\usepackage{graphicx}
\usepackage{graphics}
\usepackage[T1]{fontenc}
\usepackage[utf8]{inputenc}
\usepackage{latexsym}
\usepackage{rotating}
\usepackage{cite}
\usepackage[perpage,marginal]{footmisc}
\usepackage[colorlinks=true]{hyperref}
\usepackage[all]{hypcap}
\usepackage{xspace} 
\usepackage[usenames]{color}
\usepackage{mathrsfs}
\usepackage{microtype}

\graphicspath{{fig/}}

\makeatletter
\renewcommand\footnoterule{%
  \kern-3\p@
  \hrule\@width2.5cm
  \kern2.6\p@}
\makeatother

\definecolor {darkgreen}{rgb}{0.2,0.7,0.2}

\newcommand\be{\begin{equation}}
\newcommand\ba{\begin{eqnarray}}
\newcommand\ee{\end{equation}}
\newcommand\ea{\end{eqnarray}}

\newcommand\bw{\begin{widetext}}
\newcommand\ew{\end{widetext}}

\newcommand{\mrm}{\mathrm}

\begin{document}
\title{Binary Love Relations} 

\author{Kent Yagi}
\address{Department of Physics, Princeton University, Princeton, New Jersey 08544, USA}
\address{eXtreme Gravity Institute, Department of Physics, Montana State University, Bozeman, Montana 59717, USA}

\author{Nicol\'as Yunes}
\address{eXtreme Gravity Institute, Department of Physics, Montana State University, Bozeman, Montana 59717, USA}


\date{\today}

\begin{abstract} 

When in a tight binary, the mutual tidal deformations of neutron stars get imprinted onto observables, encoding information about their internal structure at supranuclear densities and gravity in the extreme-gravity regime. 
Gravitational wave observations of their late binary inspiral may serve as a tool to extract the individual tidal deformabilities, but this is made difficult by degeneracies between them in the gravitational wave model.   
We here resolve this problem by discovering approximately equation-of-state-insensitive relations between dimensionless combinations of the individual tidal deformabilities.  
We show that these relations break degeneracies in the gravitational wave model, allowing for the accurate extraction of both deformabilities. 
Such measurements can be used to better differentiate between equation-of-state models, and improve tests of General Relativity and cosmology. 

\end{abstract}

\pacs{04.30.Db,04.50Kd,04.25.Nx,97.60.Jd}


\maketitle

\allowdisplaybreaks{}

\emph{Tidally Deformed Neutron Stars.}~Radio and X-ray observations of neutron stars (NSs) have revealed invaluable information about the properties of matter above nuclear saturation densities~\cite{lattimer_prakash2001,lattimer-prakash-review} and about the nature of gravity in the strong-field regime~\cite{will-living,lrr-2003-5} through mass and radius measurements. The era of gravitational wave (GW) astrophysics has now begun with the historic detection of signals from black hole binaries by advanced LIGO (aLIGO)~\cite{Abbott:2016blz}.
GW observations will soon reveal the late inspiral of binary NSs, which will deepen our understanding of General Relativity (GR) in the extreme-gravity regime ~\cite{lrr-2013-9} and provide more information about the NS equation of state (EoS)~\cite{read-markakis-shibata,flanagan-hinderer-love,hinderer-lackey-lang-read,damour-nagar-villain,lackey,lackey-kyutoku-spin-BHNS,delpozzo,read-matter,Favata:2013rwa,Yagi:2013baa,Wade:2014vqa,Lackey:2014fwa,Agathos:2015uaa}. 

When in a binary, NSs deform each other through tidal interactions that then imprint onto observables. The tidal deformation of one star forces its gravitational field to be non-spherically symmetric, which then affects the trajectory of the companion. The orbital trajectories directly fold into the timing model of binary pulsars, as well as the model of the GWs emitted in the late inspiral. The dominant tidal deformation can be quantified in terms of the electric-type, quadrupolar tidal deformability~\cite{hinderer-love,damour-nagar,binnington-poisson}, or simply the \emph{tidal deformability} or \emph{Love} number~\cite{love}: the ratio between the tidally-induced quadrupole moment and the external tidal field. 

The Love number does in principle enter both the timing and the GW models, but it is not an easy parameter to estimate because of its tiny effect in observables. The effect of tidal deformations typically scales as $(R_{A} /r_{12})^{5}$, where $R_{A}$ is the radius of the $A$th star and $r_{12}$ is the orbital separation. Tidal effects are negligible in systems that are widely separated, such as binary pulsars, but they are not negligible in the late inspiral of binary NSs. For example, the leading-order tidal term in the gravitational waveform phase of an equal-mass NS binary is given by $\sim -0.92 \, (\bar \lambda_A/300) (m/2.8M_\odot)^{5/3} (f/500\mrm{Hz})^{5/3}$~\cite{flanagan-hinderer-love,hinderer-lackey-lang-read}, where $m$ is the total mass, $f$ is the GW frequency and $\bar \lambda_A = \lambda_A/m_A^5$ with $\lambda_A$ and $m_A$ representing the tidal deformability and the mass of the $A$th star respectively. The impact of tidal effects on the GWs emitted is still small, making the Love numbers challenging to measure with second-generation detectors, such as aLIGO~\cite{ligo}. 

\emph{Binary Love Relations.}~One of the main difficulties that prevents aLIGO from measuring the individual Love numbers is that they are degenerate in the GW phase model. Only a certain combination of them can be easily measured (the \emph{chirp deformability} $\Lambda$)~\cite{Favata:2013rwa,Wade:2014vqa}, while other combinations affect the GW phase model much less because they enter at higher order in perturbation theory.

\begin{figure*}[thb]
\begin{center}
\includegraphics[width=7.3cm,clip=true]{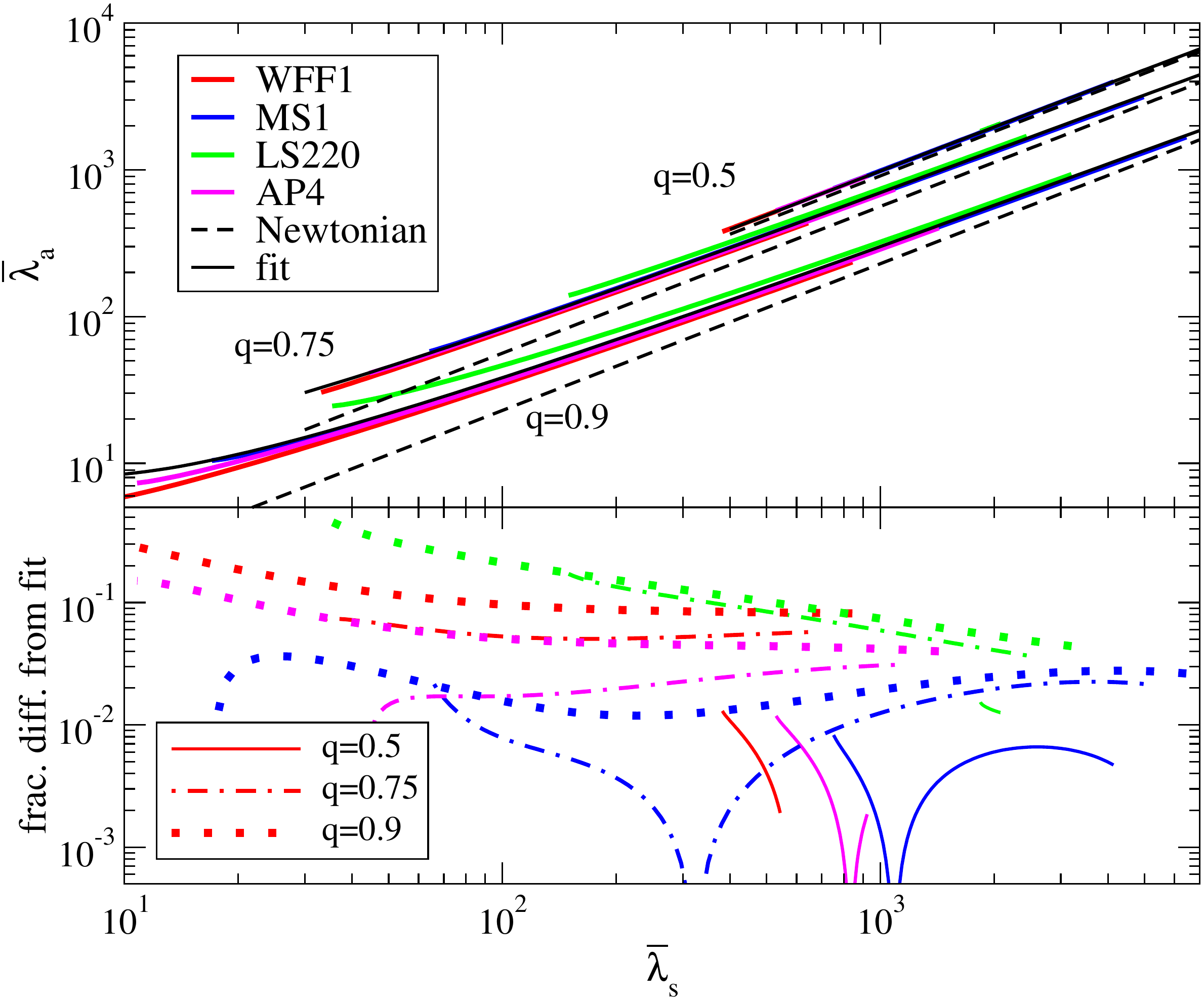}  
\includegraphics[width=7.5cm,clip=true]{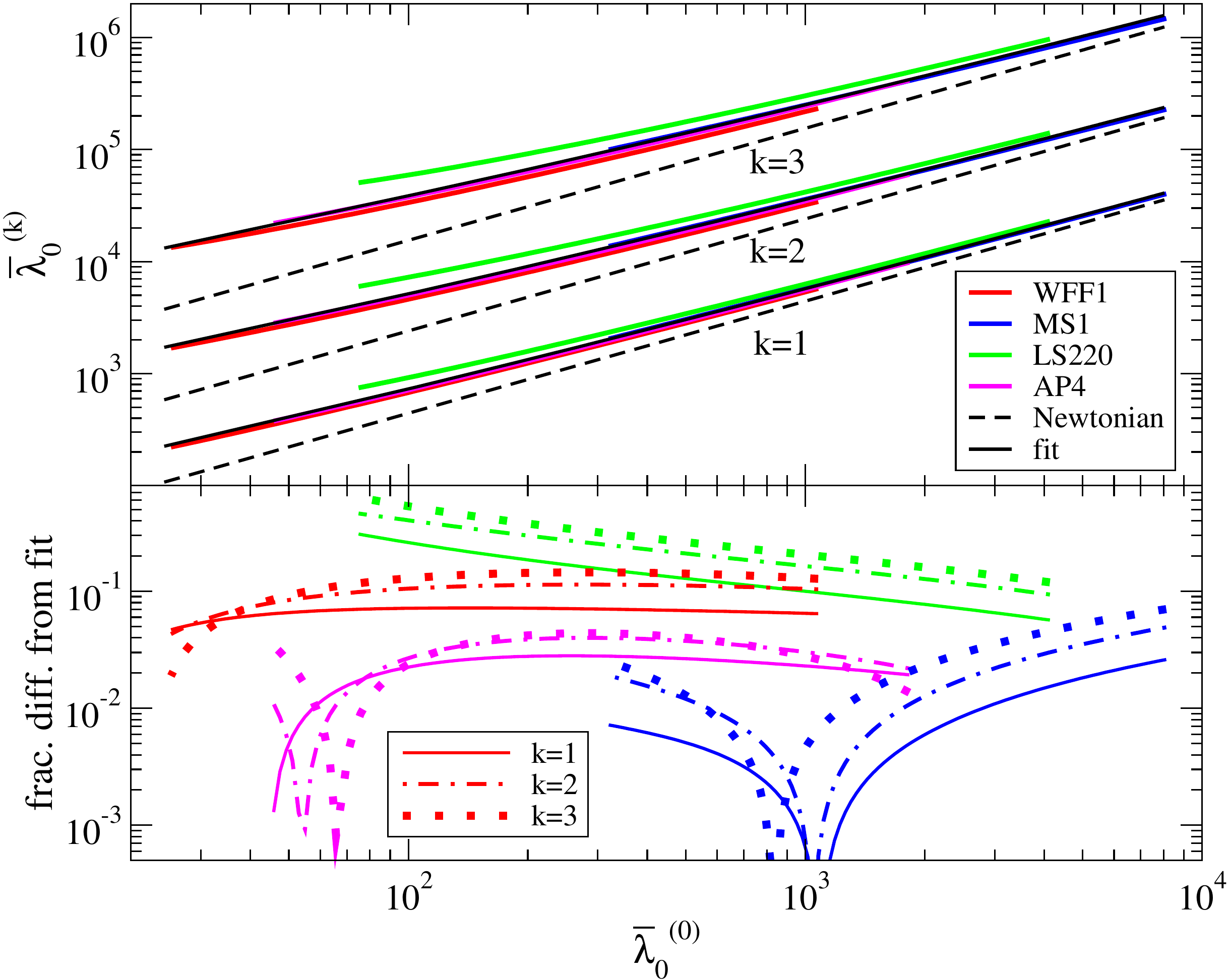}  
\caption{\label{fig:lambdas-lambdaa} (Color online) 
The $\bar \lambda_s$--$\bar \lambda_a$ relation for binaries with mass ratio $q = 0.5$, $0.75$ and $0.9$ (top left) and the $\bar \lambda_0^{(0)}$--$\bar \lambda_0^{(k)}$ relations (top right) for NSs constructed with four realistic EoSs: WFF1~\cite{Wiringa:1988tp} (red), MS1~\cite{Mueller:1996pm} (blue), LS220~\cite{LS} (green) and AP4~\cite{APR} (magenta). The black solid lines are fits to the data, while the dashed line is the Newtonian-limit (large $\bar \lambda$ limit) of these relations using a given polytropic EoS. The single parameter along each curve is $m_1$ (left) and $m_0$ (right). (Bottom) Absolute value of the fractional difference between the numerical data and the fits.} 
\end{center}
\end{figure*}

We here resolve this problem through the discovery of approximately universal (EoS insensitive) relations between the individual Love numbers of non-rotating\footnote{For stationary tidal perturbations, the leading spin correction to $\lambda$ enters at quadratic order~\cite{Pani:2015nua}, while the fastest spinning pulsar observed in a binary NS has a dimensionless spin of $\chi \lesssim 0.03$~\cite{damour-nagar-villain}. Thus, NS spins affect $\lambda$ by less than 0.1\%, which is much smaller than the EoS variation in the universal relations presented in Fig.~\ref{fig:lambdas-lambdaa}, and hence, such an effect is negligible.} NSs in a binary system. Such relations, in fact, also improve our ability to probe strong-field gravity and cosmology with GWs from NS binaries. We calculate the Love numbers of isolated stars for various masses and EoSs treating the tidal effects as small perturbations, and then solving the perturbed Einstein equations~\cite{hinderer-love,damour-nagar,binnington-poisson}\footnote{The leading nonlinear tidal effect in the gravitational waveform phase is $\sim 2.2\times 10^{-3} \, (\bar \lambda_A/300)^2 (m/2.8M_\odot)^{5} (f/1000\mrm{Hz})^{5}$~\cite{Yagi:2013sva}, and hence it is negligible.}. We extract the tidally-induced quadrupole moment and the external tidal field from the asymptotic behavior of the gravitational potential at spatial infinity, and then take the ratio of the two to find the Love number. The individual Love numbers allow us to construct a (dimensionless) symmetric combination, $\bar{\lambda}_{s} = (\bar{\lambda}_{1} + \bar{\lambda}_{2})/2$, and an antisymmetric combination, $\bar{\lambda}_{a} = (\bar{\lambda}_{1} - \bar{\lambda}_{2})/2$, where $\bar \lambda_A = \lambda_A/m_A^5$. Since the Love numbers are functions of the NS mass,  $\lambda_{A} = \lambda_{A}(m_{A})$, one can also Taylor expand the Love number about a fiducial mass $m_{0}$ through $\bar \lambda_A (m_A) = \sum_k \bar \lambda_0^{(k)} \left(1 - m_A/m_0  \right)^k /k!$ and then extract the (dimensionless) $k$th coefficient of the Taylor expansion $\bar{\lambda}_{0}^{(k)}$~\cite{damour-nagar-villain,messenger-read,delpozzo,Agathos:2015uaa}.

The left panel of Fig.~\ref{fig:lambdas-lambdaa} shows $\bar{\lambda}_{a}$ as a function of the $\bar{\lambda}_{s}$ for binaries with different mass ratio and NSs constructed with four different realistic EoSs that cover a wide range of soft to stiff EoSs. One may naively expect the relation between any two NS physical quantities to depend strongly on the underlying EoSs, as exemplified by the famous mass-radius relation~\cite{lattimer_prakash2001,lattimer-prakash-review}. However, observe that for a given value of $q := m_{1}/m_{2} < 1$, the $\bar \lambda_s$--$\bar \lambda_a$ relation is approximately EoS independent when $q$ is small \emph{or} when $\bar \lambda_s$ is large (small $m_1$). The single parameter along each curve is $m_1$, where we assume $m_{1} \in (1 M_{\odot}, m_{1}^{\rm max})$, where $m_{1}^{\rm max}$ is the maximum mass allowed for a stable NS given the particular EoS. The approximate universality remains for a wider class of realistic EoSs~\cite{longer-version}, all of which have $m_{1}^{\rm max} > 2 M_{\odot}$ to allow for the recently observed $2 M_{\odot}$ star~\cite{2.01NS}. We have also fitted the data to a Pad\'e function with a controlling factor determined from the low-compactness (``Newtonian'') limit of the $\bar \lambda_s$--$\bar \lambda_a$ relation~\cite{longer-version}:
\be
\label{eq:fit1}
\bar \lambda_a = F_{\bar n} (q) \; \bar \lambda_s \; \frac{a + \sum_{i=1}^3 \sum_{j=1}^2 b_{ij} q^j \bar \lambda_s^{-i/5} }{a + \sum_{i=1}^3 \sum_{j=1}^2 c_{ij} q^j \bar \lambda_s^{-i/5} }\,, 
\quad 
F_{n} (q) \equiv \frac{1-q^{10/(3-n)}}{1+q^{10/(3-n)}}\,,
\ee
with fitting parameters given in Table~\ref{Table:fit1}. $\bar n = 0.743$ corresponds to the average of effective polytropic indices for example EoSs considered in this letter. The bottom left panel of Fig.~\ref{fig:lambdas-lambdaa} shows the relative fractional difference between the data and the Pad\'e fit; the relations are universal to $\mathcal{O}(10\%)$, deteriorating in the equal-mass or the large mass limit\footnote{The fractional difference vanishes in certain cases, which is an artifact of the fit crossing the numerical relation with a fixed EoS and $q$ or $k$. Although the location of these minimal differences changes if one uses different fits, this is irrelevant when studying the overall EoS variation of the relations.}. 

\fulltable{\label{Table:fit1} Fitting parameters for the $\bar \lambda_s$--$\bar \lambda_a$ relation in Eq.~\eqref{eq:fit1}.
\vspace{3mm}}
\hline
\hline
\noalign{\smallskip}
\multicolumn{1}{c}{$a$}
& \multicolumn{1}{c}{$b_{11}$} 
& \multicolumn{1}{c}{$b_{12}$} 
& \multicolumn{1}{c}{$b_{21}$} 
& \multicolumn{1}{c}{$b_{22}$} 
& \multicolumn{1}{c}{$b_{31}$} 
& \multicolumn{1}{c}{$b_{32}$} 
\\
\hline
\noalign{\smallskip}
\multicolumn{1}{c}{0.07550} & 
\multicolumn{1}{c}{$-2.235$} & 
\multicolumn{1}{c}{0.8474} &
\multicolumn{1}{c}{10.45} & 
\multicolumn{1}{c}{$-3.251$} & 
\multicolumn{1}{c}{$-15.70$} & 
\multicolumn{1}{c}{13.61} & 
\\
\noalign{\smallskip}
\multicolumn{1}{c}{$c_{11}$}
& \multicolumn{1}{c}{$c_{12}$} 
& \multicolumn{1}{c}{$c_{21}$} 
& \multicolumn{1}{c}{$c_{22}$} 
& \multicolumn{1}{c}{$c_{31}$} 
& \multicolumn{1}{c}{$c_{32}$} 
& \multicolumn{1}{c}{} 
\\
\hline
\noalign{\smallskip}
\multicolumn{1}{c}{$-2.048$} & 
\multicolumn{1}{c}{$0.5976$} & 
\multicolumn{1}{c}{7.941} &
\multicolumn{1}{c}{0.5658} & 
\multicolumn{1}{c}{$-7.360$} & 
\multicolumn{1}{c}{$-1.320 $} & 
\multicolumn{1}{c}{} & 
 \\
\noalign{\smallskip}
\hline
\hline
\endfulltable

Let us discuss why the $\bar \lambda_s$--$\bar \lambda_a$ relation is universal to $\mathcal{O}(10\%)$. For Newtonian polytropes with a polytropic index $n$, the stellar mass and tidal deformability is given by $m_A \propto C_A^{(3-n)/2}$ and $\bar \lambda_A  \propto C_A^{-5}$~\cite{I-Love-Q-PRD}, where $C_A$ is the compactness of the $A$th star, i.e.~the ratio of the mass to the radius. One can easily derive the $\bar \lambda_s$--$\bar \lambda_a$ relation $\bar \lambda_a = F_n(q) \bar \lambda_s$ with $F_n(q)$ defined in Eq.~\eqref{eq:fit1}. The relation becomes exactly EoS insensitive when $q \to 0$ ($F_n \to 1$), while the EoS variation is maximum when $q \to 1$. In the latter case, the fractional difference in the relation using a polytrope of index $n$ and another polytrope with an averaged index $\bar n$ is $ [F_n(1)-F_{\bar n}(1)]/F_{\bar n}(1) = (n-\bar n)/(3 - \bar n) \lesssim 0.1$, where we used that realistic NS EoSs have $n \in (0.5,1)$~\cite{lattimer_prakash2001,flanagan-hinderer-love}. Therefore, the reason the relation is approximately universal is because (i) the relation only depends on the polytropic index and not on the polytropic constant, and (ii) the polytropic index for realistic EoSs lie in a rather narrow range.

The right panel of Fig.~\ref{fig:lambdas-lambdaa} shows $\bar{\lambda}_{0}^{(k)}$ as a function of $\bar{\lambda}_{0}^{(0)}$ for NSs constructed with a variety of EoSs. The single parameter along each curve is $m_0$. Observe that these relations are approximately universal for small $k$ and large $\bar \lambda_0^{(0)}$. We have also fitted the data to a polynomial function of $\bar{\lambda}_{0}^{(0)}$ with a controlling factor determined by the Newtonian $\bar \lambda_0^{(0)}$--$\bar \lambda_0^{(k)}$ relations~\cite{longer-version}, namely
\be
\label{eq:fit2}
\bar \lambda_0^{(k)} = \frac{\Gamma \left( k + \frac{10}{3-\bar n} \right)}{\Gamma \left(\frac{10}{3- \bar n} \right)} \ \bar \lambda_0^{(0)} \left( 1 + \sum_{i=1}^3 a_{i,k} \ \bar \lambda_0^{(0)}{}^{-i/5} \right)\,, 
\ee
where $\Gamma (x)$ is the Gamma function and fitting parameters are given in Table~\ref{Table:fit2}. The bottom right panel of Fig.~\ref{fig:lambdas-lambdaa} shows the relative fractional difference between the data and the polynomial fit; these relations are also universal to $\mathcal{O}(10\%)$, deteriorating as $k$ increases.

\fulltable{\label{Table:fit2} Fitting parameters for the $\bar \lambda_0^{(0)}$--$\bar \lambda_0^{(k)}$ relation in Eq.~\eqref{eq:fit2}.
\vspace{3mm}}
\hline
\hline
\noalign{\smallskip}
\multicolumn{1}{c}{$a_{1,1}$}
& \multicolumn{1}{c}{$a_{2,1}$} 
& \multicolumn{1}{c}{$a_{3,1}$} 
& \multicolumn{1}{c}{$a_{1,2}$} 
& \multicolumn{1}{c}{$a_{2,2}$} 
& \multicolumn{1}{c}{$a_{3,2}$} 
& \multicolumn{1}{c}{$a_{1,3}$} 
& \multicolumn{1}{c}{$a_{2,3}$} 
& \multicolumn{1}{c}{$a_{3,3}$} 
\\
\hline
\noalign{\smallskip}
\multicolumn{1}{c}{0.4443} & 
\multicolumn{1}{c}{2.726} & 
\multicolumn{1}{c}{$-0.6350$} &
\multicolumn{1}{c}{0.3344} & 
\multicolumn{1}{c}{6.568} & 
\multicolumn{1}{c}{$-0.4671$} & 
\multicolumn{1}{c}{$-0.1334$} & 
\multicolumn{1}{c}{11.35} & 
\multicolumn{1}{c}{$-3.928$} & 
\\
\noalign{\smallskip}
\hline
\hline
\endfulltable

\emph{Applications to GW Astrophysics.}~The binary Love relations (either the $\bar \lambda_s$--$\bar \lambda_a$ or the $\bar \lambda_0^{(0)}$--$\bar \lambda_0^{(k)}$ one~\footnote{Strictly speaking, the $\bar \lambda_0^{(0)}$--$\bar \lambda_0^{(k)}$ relation holds for a single NS and does not require two NSs.}) can be used to break degeneracies between the individual tidal deformabilities $\bar \lambda_{1,2}$ that enter the GW model. For example, re-expressing $\bar \lambda_{a}=\bar \lambda_{a}(q,\bar \lambda_{s})$ in the GW phase removes $\bar \lambda_{a}$ from the parameter list in the GW model, thus improving the accuracy to which $\bar \lambda_{s}$ can be measured. Once $\bar \lambda_{s}$ has been estimated, one can use the $\bar \lambda_s$--$\bar \lambda_a$ relation again to find $\bar \lambda_{a}$, and from knowledge of both $\bar \lambda_{s}$ and $\bar \lambda_{a}$ one can find the individual deformabilities $\bar \lambda_{1}$ and $\bar \lambda_{2}$.    

\begin{figure}[thb]
\begin{center}
\includegraphics[width=7.5cm,clip=true]{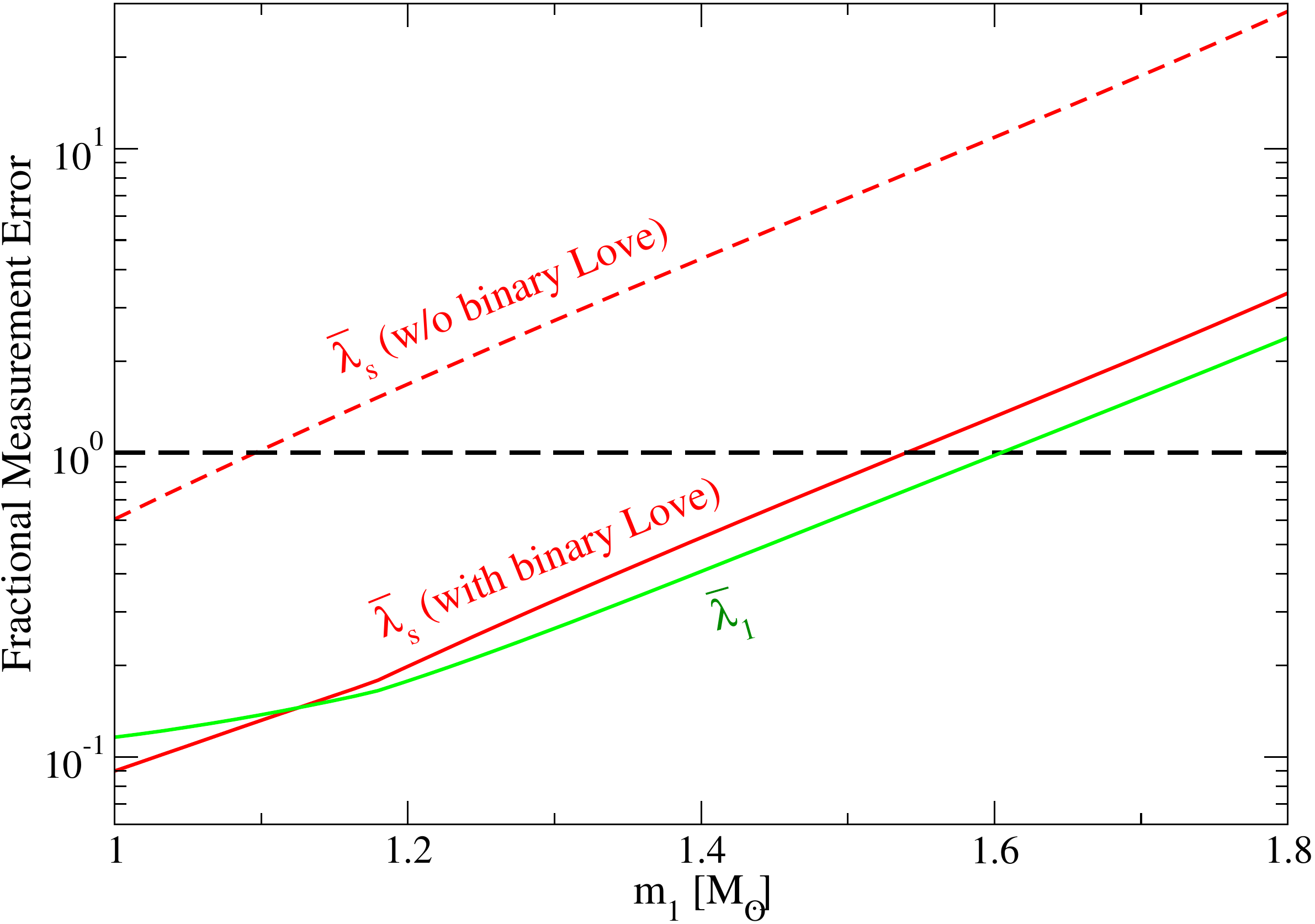}  
\caption{\label{fig:Fisher-prior} (Color online) Estimated fractional measurement accuracy of $\bar \lambda_s$ with (solid red) and without (dashed red) using the binary Love relation, given GW observations of NS binaries with aLIGO as a function of $m_{1}$ with $q = 0.9$. The figure was generated through a Fisher analysis that assumed a signal-to-noise ratio (SNR) of 30, that both NSs are constructed with the AP4 EoS~\cite{APR}, and the zero-detuned aLIGO spectral noise density curve~\cite{ajith-zero-detuned}. We further show the fractional measurement accuracy of the individual tidal deformability $\bar \lambda_1$. Observe that the binary Love relation improves the measurement accuracy by approximately an order of magnitude.  
}
\end{center}
\end{figure}

Figure~\ref{fig:Fisher-prior} shows an estimate of the statistical fractional measurement accuracy of $\bar{\lambda}_{s}$ with and without the binary Love relation, as a function of the mass of one of the NSs assuming aLIGO detections. Such a statistical accuracy is obtained from a Fisher analysis with an SNR of 30\footnote{From the expected detection rate of NS binaries with aLIGO in~\cite{abadie}, one finds that the expected rate with SNR=30 to be $\sim 1/$yr. Moreover, such an SNR is similar to that of GW150914~\cite{Abbott:2016blz}.}, taking correlations with all search parameters into account. Observe that the accuracy to which $\bar{\lambda}_{s}$ can be measured increases by roughly an order of magnitude when one uses the binary Love relations. Figure~\ref{fig:Fisher-prior} also shows the accuracy to which $\bar{\lambda}_{1}$ can be measured through propagation of errors. Observe that the individual deformabilities can be estimated for NS binaries with masses less than approximately $1.6 M_{\odot}$. Moreover, the fractional measurement accuracy of $\bar \lambda_1$ using the binary Love relation is better than that of the chirp tidal deformability $\Lambda$ by $\sim 50\%$, as we will show elsewhere~\cite{longer-version}. 

The binary Love relations, however, are not exact and thus the EoS variation in the relations introduces a systematic error in the determination of the tidal deformabilities. This error, however, is smaller than $5\%$ for all cases considered~\cite{longer-version}, and thus much smaller than the statistical error presented in Fig.~\ref{fig:Fisher-prior}. This is because the term that depends on $\bar \lambda_a$ in the GW model is suppressed by the ratio $(m_2-m_1)/(m_1 + m_2)$, so the systematic error due to the approximate nature of the universality is suppressed precisely in the region where it is largest.   

\begin{figure}[thb]
\begin{center}
\includegraphics[width=7.cm,clip=true]{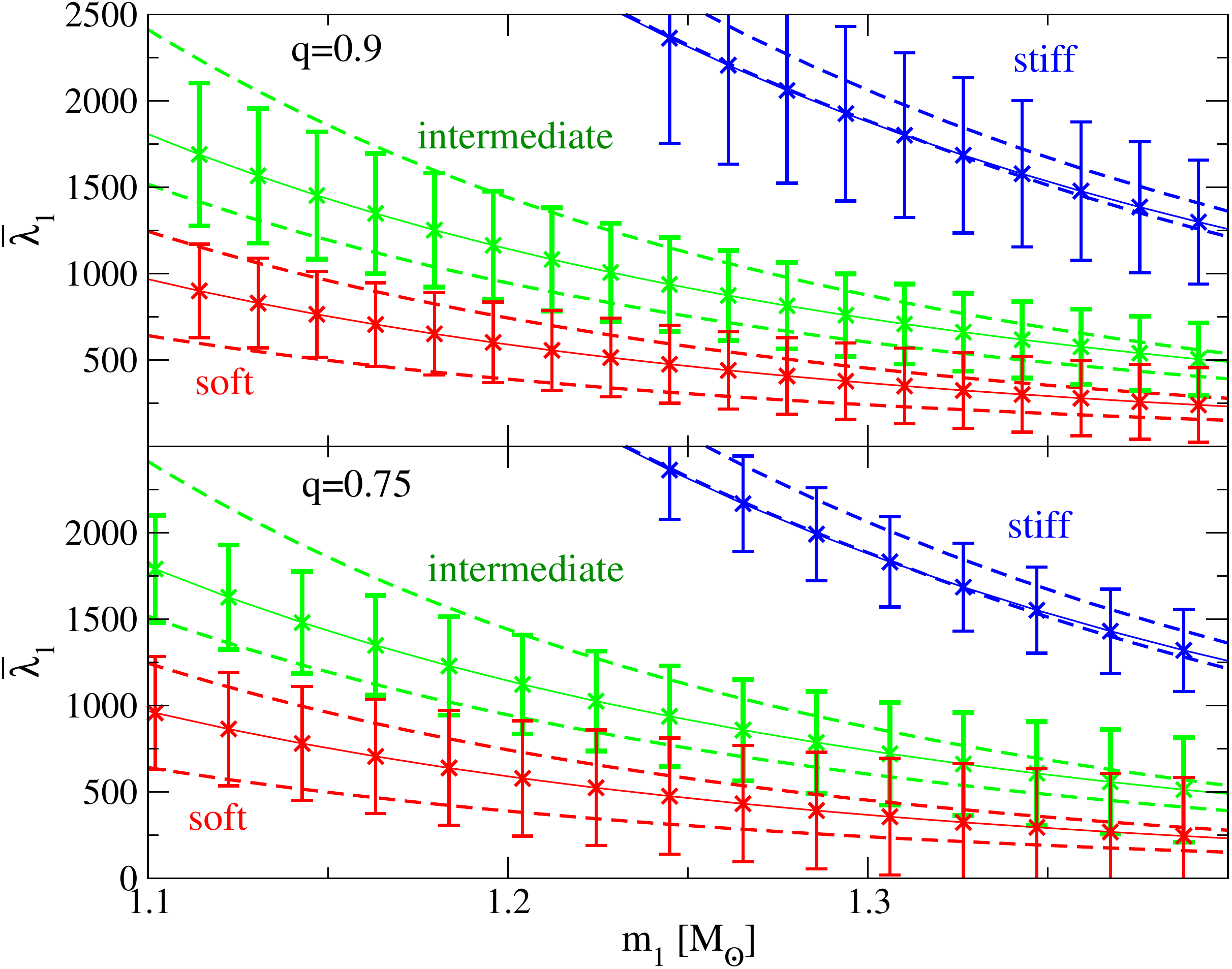}  
\includegraphics[width=6.8cm,clip=true]{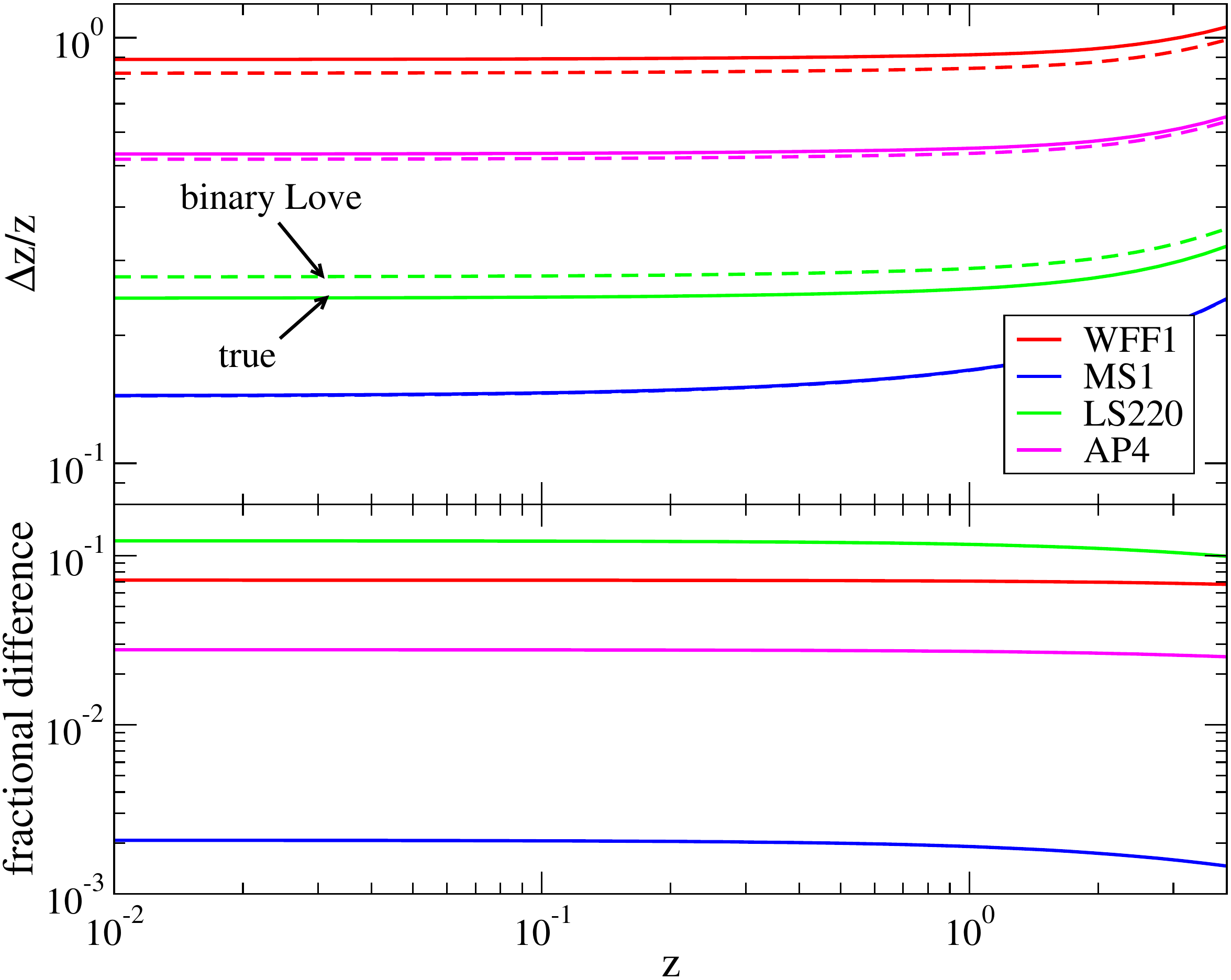}  
\caption{\label{fig:lambda-m-error} (Color Online) (Left) The $\bar{\lambda}_{1}$--$m_{1}$ relation for three different EoS classes: soft (red dashed region), intermediate (green dashed region) and stiff (blue dashed region), together with 2-$\sigma$ error bars for the measurement of $\bar{\lambda}_{1}$ estimated through  error propagation from $\bar{\lambda}_{s}$ and $\bar{\lambda}_{a}$. We assume aLIGO detections of GWs from NS binaries of a given $m_1$, a fixed $q=0.9$ (top) and $q=0.75$ (bottom), an SNR of 30 and EoSs WFF2, MPA1 and MS1b to represent the soft, intermediate and stiff classes. Observe that the stiff class can be distinguished easily, while distinguishing the other two would require at least a low-mass NS binary observation. (Top Right) Fractional measurement accuracy of the redshift $z$ as a function of $z$ for various EoSs assuming that we know the true $\bar \lambda_0^{(1)}$ \emph{a priori} (solid), and with $\bar \lambda_0^{(1)}$ obtained from the $\bar \lambda_0^{(0)}$--$\bar \lambda_0^{(1)}$ relation (dashed). In both cases, we assume the detection of GWs emitted from $(1.4,1.4)M_\odot$ NS binaries using ET.  (Bottom Right) Fractional difference between $\Delta z/z$ with the true $\bar \lambda_0^{(1)}$ and with that obtained from the $\bar \lambda_0^{(0)}$--$\bar \lambda_0^{(1)}$ relation. The difference is at most $\sim 10\%$, so the binary Love relations can be used in place of requiring absolute knowledge of the EoS \emph{a priori}.}
\end{center}
\end{figure}

\emph{Applications to Nuclear Physics.}~The relation between the tidal deformability of a given NS and its mass encodes the NS EoS. The left panel of Fig.~\ref{fig:lambda-m-error} shows the $\bar{\lambda}_{1}$--$m_{1}$ relation for three different classes of EoSs: a \emph{stiff family} (Shen~\cite{Shen1,Shen2}, MS1~\cite{Mueller:1996pm} and MS1b~\cite{Mueller:1996pm}); an \emph{intermediate family} (AP3~\cite{APR}, MPA1~\cite{1987PhLB..199..469M}, LS220~\cite{LS} and ENG~\cite{1996ApJ...469..794E}); and a \emph{soft family} (WFF1~\cite{Wiringa:1988tp}, WFF2~\cite{Wiringa:1988tp}, SLy~\cite{SLy}, AP4~\cite{APR}). These example EoSs were chosen such that they cover a wide range of stiffnesses and can support the recently observed $2 M_{\odot}$ NS~\cite{2.01NS}.

The measurement of the individual tidal deformabilities thanks to the binary Love relations allow us to constrain EoS families given a GW observation. Figure~\ref{fig:lambda-m-error} shows the accuracy to which $\bar{\lambda}_{1}$ could be measured using the binary Love relations, assuming aLIGO observations of NSs with different $m_{1}$ and fixed mass ratio $q = 0.9$ (top) and $q=0.75$ (bottom). Observe that the stiff family can always be distinguished from the soft and intermediate families. Observe further that the latter two may also be separated if aLIGO detects sufficiently light NSs, though this depends strongly on the choice of fiducial EoS within each class. Moreover, we found that distinguishing the soft and intermediate families becomes more difficult as the mass ratio becomes smaller. A similar plot can be obtained using $\Lambda$ in the GW model without the binary Love relation, but the error bars are larger~\cite{longer-version}.

\emph{Applications to Experimental Relativity.}~GR has passed a variety of tests in the Solar System and using binary pulsars~\cite{will-living,lrr-2003-5}, however it remains to be stringently constrained in the extreme-gravity regime, where gravity is non-linear and dynamical~\cite{lrr-2013-9}. Recently, we proposed the use of the approximately universal I-Love-Q relations to carry out a model-independent and EoS-insensitive test of Einstein's theory~\cite{I-Love-Q-Science,I-Love-Q-PRD}. The I-Love relation, for example, establishes that the dimensionless moment of inertia $\bar{I}$ and the Love number satisfy a relation that is approximately EoS independent, similar to the binary Love ones. This relation, however, is sensitive to the nature of the gravitational interaction in the extreme-gravity regime, and thus, the $\bar{I}$--$\bar{\lambda}$ relation is different in GR and in certain modified gravity theories. The independent measurements of $\bar{I}$ through radio pulsar observations and of $\bar{\lambda}$ through GW observations could allow us to verify the GR $\bar{I}$--$\bar{\lambda}$ relation and constrain deviations in a model-independent way. 

For such a test to be effective, however, GW observations would have to be able to measure the tidal deformability of the $A$th NS $\bar{\lambda}_{A}$ to sufficient accuracy. In~\cite{I-Love-Q-Science,I-Love-Q-PRD}, we considered an equal-mass NS binary, but we have here showed that through the binary Love relations one can infer an estimate of the individual Love numbers, and thus carry out this test, even for unequal-mass binaries. Alternatively, one can use the $\bar{I}$--$\bar{\lambda}$ relation by measuring $\bar \lambda_0^{(0)}$ with $m_0$ set equal to the mass of the NS for which $\bar I$ is measured. The $\bar \lambda_0^{(0)}$--$\bar \lambda_0^{(k)}$ relations improve the ability of testing strong-field gravity, since such relations improve the measurement accuracy of $\bar \lambda_0^{(0)}$. 

Precisely how this improvement in the I-Love test translates into an improved constraints on the coupling constants of the modified theory depends sensitively on the particular class of theory considered. For example, in dynamical Chern-Simons (dCS) gravity~\cite{jackiw,Smith:2007jm,CSreview}, a theory that has not yet been stringently constrained with Solar System and binary pulsar observations, the binary Love relations improve the I-Love constraints by approximately $5\%$. This is because constraints on the dCS coupling parameter are dominated by the smallest scale of the problem, which is here the NS radius, and are much less sensitive to the accuracy to which $\bar{I}$ and $\bar{\lambda}$ can be measured. 

\emph{Applications to Cosmology.}~The tidal deformabilities of NSs depend on their intrinsic mass for a given EoS (the Love-m relation), while the GW model depends both on the tidal deformabilities and on the redshifted masses for sources at cosmological distances. Therefore, if a GW observation can be used to extract the redshifted masses and the Love numbers of a given NS binary, one can then use the Love-m relation to extract the intrinsic masses and from that the binary's redshift~\cite{messenger-read,Li:2013via}. With this at hand, one can then use the GW amplitude to extract the luminosity distance, and from this estimate cosmological parameters, such as the Hubble parameter today and the energy densities of matter and dark energy~\cite{DelPozzo:2015bna}. 

This application of GW astrophysics to cosmology relies sensitively on knowing the Love-m relation. Indeed, it is this relation that provides an estimate for the intrinsic masses of the binary, given the Love number. But the Love-m relation depends on the EoS, and thus, this application relies on the assumption that we know the correct EoS \emph{a priori}. Such a stringent assumption can be lifted by using the $\bar \lambda_0^{(0)}$--$\bar \lambda_0^{(k)}$ relation, since this can be used to map the Love-m relation and obtain the intrinsic masses.

The accuracy to which the cosmological parameters can be determined depends on the accuracy to which the redshift can be extracted. The latter is only sufficiently good when one uses observations from third-generation detectors, such as the Einstein Telescope (ET). The right panel of Fig.~\ref{fig:lambda-m-error} shows the measurement accuracy of the redshift using the binary Love relations or assuming the EoS is known exactly \emph{a priori} for a set of sources at different redshift and with different EoSs. This accuracy was estimated through a Fisher analysis similar to that used in Fig.~\ref{fig:Fisher-prior}, but with the tidal deformabilities replaced by redshift in the parameter set~\cite{messenger-read}. Observe that the binary Love relations affect the accuracy to which the redshift can be determined only by at most $10\%$. This means that we can still consider this GW application to cosmology without assuming exact knowledge of the EoS thanks to the binary Love relations. 

\emph{Future Directions.}~The parameter estimation analysis presented here is a proof-of-principle of the applicability of the binary Love relations to GW astrophysics, nuclear physics, experimental relativity and GW cosmology. These results could be strengthened by carrying out a Bayesian and model selection study. Ultimately, once aLIGO detects GWs from NS binaries, the binary Love relations could be folded into the GW models currently employed to improve the accuracy to which the deformabilities can be extracted.  

Future work could also focus on understanding whether other approximately universal relations exist between higher-order tidal parameters. Indeed, the electric-type, quadrupole tidal deformability is only the first in an infinite series of deformabilities that enter at ever higher order in perturbation theory. Given the universality found here for a binary system and the universal relations among higher-order tidal parameters of isolated NSs found in~\cite{Yagi:2013sva}, similar approximately universal relations may exist among higher-order tidal parameters. Such putative higher-order relations seem irrelevant for second-generation GW interferometers such as aLIGO, but may be relevant for third generation detectors such as LIGO III and Einstein Telescope~\cite{Yagi:2013sva}.


\emph{Acknowledgments}.~We thank Katerina Chatziioannou for comments and suggestions. K.Y.~acknowledges support from JSPS Postdoctoral Fellowships for Research Abroad and the NSF grant PHY-1305682, while N.Y. acknowledges support from the NSF CAREER Grant PHY-1250636. Some calculations used the computer algebra-systems MAPLE, in combination with the GRTENSORII package~\cite{grtensor}.

\section*{References}

\bibliographystyle{iopart-num}
\bibliography{master}
\end{document}